ID: A-2073

# A review on multiscale computational studies for enhanced oil recovery using nanoparticles

**Rajneesh Kashyap**, **Mohit Kalra, Arti Kashyap**
School of Basic Sciences, Indian Institute of Technology Mandi, H.P.-175005, India

***Presenting author, E-mail address: rajneeshkashyap.rk13@gmail.com***

## Abstract
Oil reservoirs around the globe are at their declining phase and in spite of enormous effectiveness of Enhanced Oil Recovery(EOR) in the Tertiary Stage. This process still bypasses some oil reason being surface forces responsible for holding oil inside the rock surface which are not being altered by the application of existing technologies. The processes coming under Tertiary Section Supplements primary and secondary sections. However, the mechanism of operating is different in both. Nanoparticles are showing a significant role in EOR techniques and is a promising approach to increase crude oil extraction. This is due to the fact that size of nanoparticles used for EOR lies in the range of 1-100 nm. It is also an interesting fact that in different operational conditions and parameters, the performance of nanoparticles also vary and some are more effective than others, which leads to various levels of recovery in the EOR process. In the present study, we intend to summarize a report having an up to date status on nanotechnology assisted EOR mechanisms where nanoparticles are used as nano-catalysts, nano-emulsions and nanoparticles assisted EOR mechanisms to destabilize the oil layer on carbonate surface. This review also highlights the various mechanisms such Gibb’s free energy, wettability alteration, and Interfacial Tension Reduction (ITR) including interaction of available nanoparticles with reservoirs. Experimental measurements for a wide range of nanoparticles are not only expensive but are challenging because of the relatively small size, especially for the measurements of thinner capillaries of a nanoscale diameter. Therefore, we considered computational simulations as a more adequate approach to gain more microscopic insights into the oil displacement process to classify the suitability of nanomaterials. Also, the computational modelling will be helpful for making decisions and cost effective planning for EOR.

## Introduction
Population and global industrial growth rate are leading to increase the demand of energy (1). In order to meet the need of energy requirement, crude oil resources which is a function of new hydrocarbon discoveries are being explored by improving the recovery mechanism of existing oil fields. Lamentably, the existing oil and gas resources are at declining phase (1). In order to meet the continuous increasing oil demand and keep sustainable oil supply, the ultimate recoverable oil reserves should be increased. This could be achieved by replacing the produced oil by new reserves from new discoveries as well as applying enhanced/improved oil recovery (EOR/IOR) methods to maximize oil recovery from current mature oil fields by minimizing the trapped oil left behind in the reservoirs. EOR/IOR methods demand no new technology and utilize existing equipment/machinery and facilities of the developed oil reservoirs (1).

Over the past few decades, the research activities and research trend has shifted in diverse disciplines, started from the “Macro-domain” and now has reached to the “Microdomain” and “Nano-domain” and probably in the coming years it will go to “Pico-domain” and “Femto-domain”. Various areas like microfluidics, nanoscience and nanotechnology and nanofluidics have observed an increase in the research area due to the advancing trend of scaling down in technological innovations. Nanoparticles are tailored to have at least a dimension in the order of 100nm. The uniqueness of nanoparticles is due to their size-dependent properties, which are result of their large surface-to-volume ratio therefore, this enables surface atoms/molecules to have notable impact on its properties (2,3). These methodologies are not only more energy efficient but also have environmental benefits which can be anticipated to play a paramount contribution in future oil supply. The last two decades have seen considerable amount of improvement in this area like wettability has been given a great interest as new water based tertiary oil recovery process (4). Various investigations, experimental studies as well as some pilot projects/studies have verified the validity and potential of this phenomenon, still, the mechanism is ill-defined. One of the main reasons for considering the wettability as EOR method is that with this technique the trapped immobile residual oil is targeted i.e. saturated oil left behind trapped after traditional water-flooding and various techniques.

The Multiscale computational approaches /models such as Density Functional Theory (DFT), Molecular Dynamics (MD) and Computational Fluid Dynamics (CFD) are widely used for EOR. Various parameters that can be optimized using these computational approaches, individually, are as follows:
**Density Functional Theory (DFT):** Atomistic and Electronic properties enhancement, band structure of nanomaterials, Gibb's free energy etc.
**Molecular Dynamics (MD):** Binding energy, wettability, equilibrium and non-equilibrium process parameter, surface properties, Inter facial tension, adsorption energy etc.
**Computational Fluid Dynamics (CFD):** Oil flow through pipelines, effect of pressure & temp, viscosity, stress, various mechanical and flow properties.
However, computational modeling in the field of crude oil extraction and recovery has not been explored widely. Efforts are very sparse and limited despite having possibility of huge impact. Considering the usefulness of computational modelling for EOR, this review article will give a clear insight and work done so far in this field.

## Application of Nanotechnology in Oil recovery

In order to supplement the existing oil recovery processes, exploration into making use of nanotechnology has come out as an embryonic alternative for the tertiary oil recovery techniques. "Nano-EOR" is the term used for describing applications of nanoparticles in Enhanced Oil Recovery (EOR) processes. Wettability alteration and interfacial tension reduction are the two mechanisms by which nanoparticles assist in improving the recovery or residual oil. The combination of these two may not be the only underlying mechanism as several other approaches have been offered for nano-EOR. This is due to the fact that size of nanoparticles used for EOR lies in the range of 1-100 nm (5). Density-Functional-Theory calculations of interaction energies reveal that due to their large surface to volume ratio they exhibit noble properties in comparison with the same bulk molecules since there is much higher concentration of atoms at their surface due to their exceptionally small size (6). Due to this fact, they can move along the narrow pores and cause massive diffusion. It is worth pointing out that utilizing Density-Functional-Theory and Quantum Molecular Dynamics, it was investigated that nanoparticles are showing remarkable contribution in minimizing oil viscosity, improving the mobility ratio, and till now no investigation has been made to improve the EOR process by altering the reservoir permeability (7). The schematic of nano-fluid wettability alteration is illustrated in **Figure 1**. It is also an interesting fact that in different operational conditions and parameters, the performance of nanoparticles also vary and some are more effective than others, which leads to various levels of recovery in the EOR process.

## Nanoparticle EOR Mechanism

**Altering Wettability and Interfacial Tension Reduction using Nanoparticles:** If a rock surface is hydrophobic, the oil phase present is strongly packed in it and oil recovery by standard water flooding technique will be ineffective (8). Therefore, in such cases, Density Function Theory (DFT) calculations have confirmed that oil recovering efficiency can be improved from oil-wet carbonate reservoirs by altering wettability using dilute surfactants(Cationic/Anionic) and electrolyte solutions(nanofluid) (8,9,10,11). As illustrated in **Figure 2**, Wettability is calculated by measuring contact angle when two immiscible liquids interface come in contact with each other at a solid surface. Reservoir rocks exhibit various wetting properties such as water-wet, oil-wet, or intermediate wet. The Quantum Molecular Dynamic (QMD) simulations have revealed that water-wet rocks have high affinity for water and water pre-eminently holds the position in tiny rock pores as well as at the surface of the formation rock. (10,11). The corresponding angle is determined with denser phase using Eq. i:

$$\boldsymbol{\sigma}_{so} - \boldsymbol{\sigma}_{sw} = \boldsymbol{\sigma}_{ow} \cos\boldsymbol{\Theta} \qquad [i]$$

Where, $\boldsymbol{\sigma}_{so}$ = IFT lying in between oil and rock surface [dynes/cm or N/m], $\boldsymbol{\sigma}_{sw}$ = IFT lying in between water and rock surface [dynes/cm or N/m], $\boldsymbol{\sigma}_{ow}$ = IFT lying in between oil and water [dynes/cm or N/m] and $\boldsymbol{\Theta}$ = tangential angle made by water phase at the water-oil-rock surface interface [in degrees]

The use of nanoparticles in EOR process has been reported in various literatures. As per our survey, spherical shaped Silica nanoparticles having diameter ranging from several to tens of nanometers are widely used in the EOR process. Son *et al.* (12) investigated the use of nanoparticle-stabilized oil/water emulsions as EOR agents in a column of silica bead having mineral oil. According to their result, there was an increase in the rate of recovery of oil of approximately 11% after water flooding. This happened due to the reason that there was a piston effect which created large pressure variance through the column and enabled the remaining oil in it to be produced. A comparative study was done by Ogolo etal. (13) in order to study the effect of eight different nanoparticle oxides (oxides of iron, tin, magnesium, zirconium, silicon, aluminum, nickel, zinc) on oil recovery under surface conditions. For preparing nanofluids the dispersion media used includes water, ethanol, diesel, and brine. The results confirmed that oxides of Aluminum and Silicon are best candidates for EOR application when the dispersion media is water and brine. Whereas, in the case when they were dispersed in ethanol, it was observed that most of the oil recovery was done by silane-treated oxide of silicon. One of the main discoveries of the study was that, the oxide of Aluminum reduced the oil viscosity whereas Silicon reduced the interfacial tension between oil and water as well as altered the wettability of rock.

Over the last few decades, the two phenomenon of interfacial tension reduction and wettability modification are widely being used as the mechanisms where nanoparticles assist in enhancing the recovery of oil from oil-bearing formations (11). Onyekonwu and Ogolo (14) studied various types of polysilicon nanoparticles i.e. hydrophilic and lipophilic (LHPN), hydrophobic and lipophilic (HLPN), and neutrally wet (NWPN) for enhancing oil recovery via wettability alteration. It was observed from the simulation results that both HLPN and NWPN were successful in enhancing oil recovery in case of water wet formations. The dispersion medium was ethanol and the process was achieved via wettability alteration and interfacial tension reduction. Whereas, LHPN lead to increase oil recovery in oil formations which are water wet.

**Disjoining pressure due to nanoparticles:** When a nanofluid solution comes in contact with an oil droplet on a solid surface, this leads to the formation of two contact lines (10). These two lines set apart by a layer of well-ordered nanoparticle structures, this leads to the formation of high disjoining pressures and forms a wedge-like shape spreading the nanoparticles before the inner line and after the outer line as shown in **Figure 2**. The formation of this well-defined wedge-like structure of nanoparticles is driven by the injection pressure of nanofluids which forces the nanoparticles to a confine region (10). This type of arrangement leads to increase in the entropy of the nanofluids since the freedom in nanopartices is more than in nanofluids which results in the excretion of disjoining pressure. Further, simulations using Density Function Theory also provide the insights in the forces behind structural disjoining pressure comprises of electrostatic repulsion, Brownian motion and Van-der Waals.

**Propagation of nanoparticles in porous media:** In most of the cases the oil field environments are under the conditions of high pressure, high temperature and salinity and the properties of rock are also not uniformly distributed (9,10). In such an environment, it is challenging to understand the adsorption and propagation mechanisms of nanoparticles. The nanoparticles propagate in a porous media with the help of various forces such as convection, diffusion, and hydrodynamic forces. The three mechanisms which are responsible for the transmission of nanoparticles in a porous medium have been cited by ShamsiJazeyi *et al*. (15). These process are filtered through a physical process, chemical stability in a solution, adsorption or retention on the surface of a medium which is porous and log jamming. The process of physical filtration is independent of nanoparticle dispersibility and linked with the particle size having size more than pore size in porous media (10), which further depends on the aspect ratio, shape, and, size distribution of the nanoparticles. Nanoparticle aggregation is often formed due to the chemical instability in nano-dispersions. This happens due to the reason that hydrophobic and Van der Waals interactions are inhibited because of polymer coatings over it. The stability of nanoparticles is affected due to salinity and presence of divalent ions. The stability and mobility studies of functionalized silica nanopartices at high temperature and high salt concentrations for EOR purposes were done by were done by Miranda *et al.* (16). In their study it was discovered that the presence of ions could actually modify the propagation behavior of nanoparticles. It was also observed that the coefficients of diffusion for nanoparticles increased with increasing the concentration of salt.

**Controlling the viscosity of injected fluids:** The concentration of nanoparticles presents in a nanofluid determine its viscosity (9). This can be further explained by the viscosity on an injected fluid increases because of nanoparticles due to mobility of adjacent fluid molecules reduce around the nanoparticles. Using CFD technique It was noted that the size of nanoparticles injected in the nanofluid also affect the shear viscosity of the injected fluids. Shah *et al.* (17) studied that an increase in the size of nanoparticles leads to an increase in the viscosity of fluids injected. The mobility of the injected fluid is determined by parameter “Mobility Factor (M)”. It is defined as the ratio of the displacing fluid to that of the displaced fluid. Mathematically it is represented as shown in Eq.ii:

$$M=\frac{\lambda i}{\lambda o} \quad [ii]$$

Where, $\lambda i$ and $\lambda o$ are the mobility of injected fluid and oil respectively. If the mobility ratio is $M \leq 1$ this means the injected fluid has mobility less than or equal to the mobility of the displaced oil ($\lambda i \leq \lambda o$). Such a situation happens when there is an increase in the concentration or increase in the size of the nanoparticles in the injected fluids.

## Challenges for Nano-EOR

One of the main challenges in using nanotechnology in the petroleum industry is related with the production of nanomaterials. Mass production of nanomaterials for its usage in this industry is far more expensive than conventional synthesis techniques. The production of nanoparticles is an expensive technique due to non-standardized approaches applied for its production (5). The large-scale production of nanoparticles by making use of conventional physio-chemical processes is economically and environmentally undesirable (3,4). Consequently, implementing Green Nanotechnology in the petroleum industry will be able to provide the cheap alternative for their synthesis and applications. Makarov *et al*. (18), studied the mechanism for synthesis of metal nanoparticles via biological route which comprised of three main steps i.e. activation phase, growth phase, and termination phase. Stability of nano-suspensions under harsh conditions also militate the application of these materials. Repulsive forces also known as steric hindrance is required to overcome the force of attraction between nanoparticles. A few studies have reported the usage of surfactants to be used in dispersing and stabilizing nanoparticles in various continuous medium (19).

## Conclusion

The main objective of this comprehensive review is to summarize the applications of nanotechnology that have attracted lot of interest recently in EOR. The number of publications and patents filed using nanomaterials has increasing significantly in the last two decades. This article attempts to summarize and classify the most applied computational methods in the EOR process and also explains their dominant mechanism. This review shows that there is a high demand of enormous computational research work to study the Nano-EOR mechanisms and finding effective solutions for resolving the challenges.

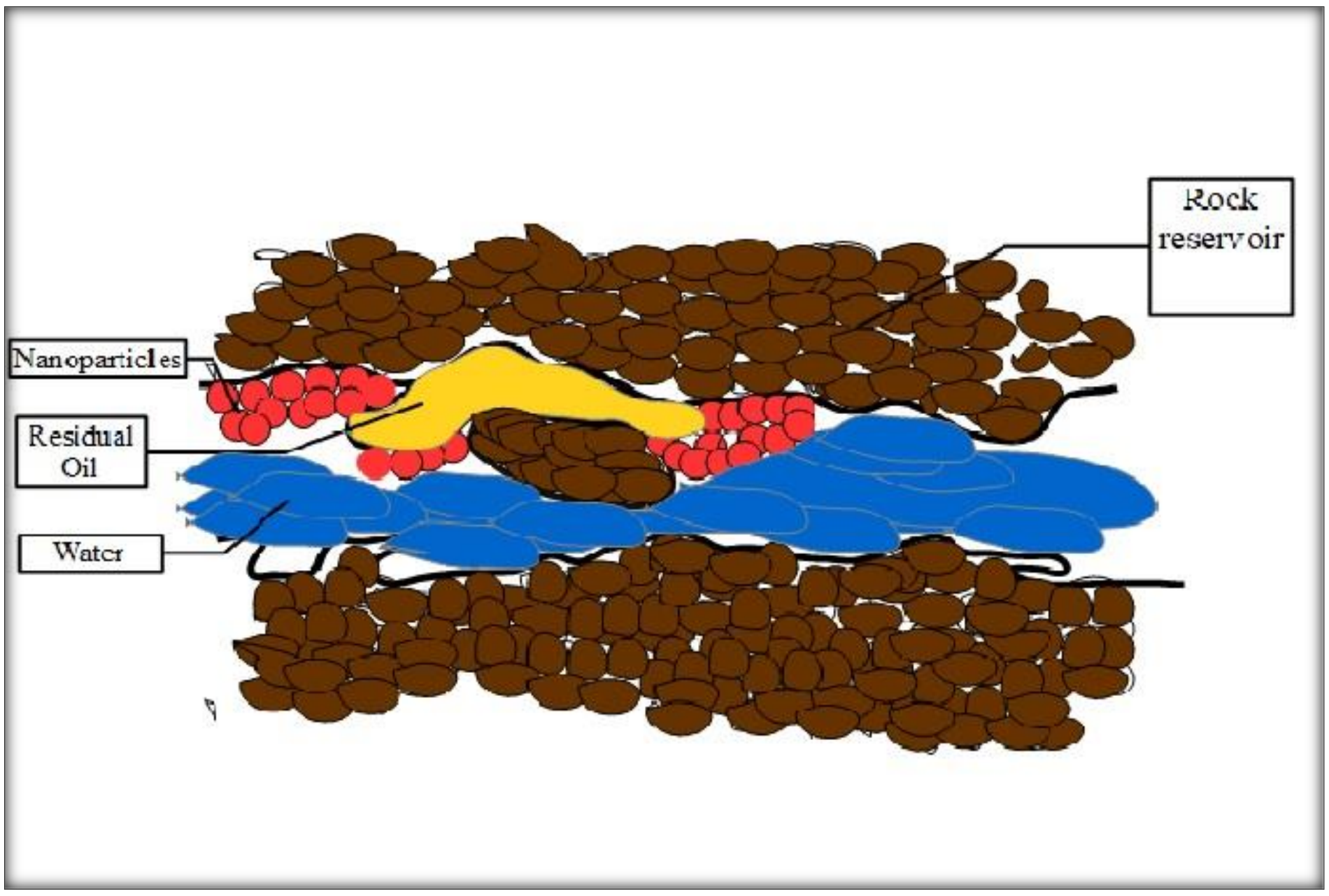

**Figure 1:** Principle of altering wettability using nanofluid, where oil is trapped in the reservoir.

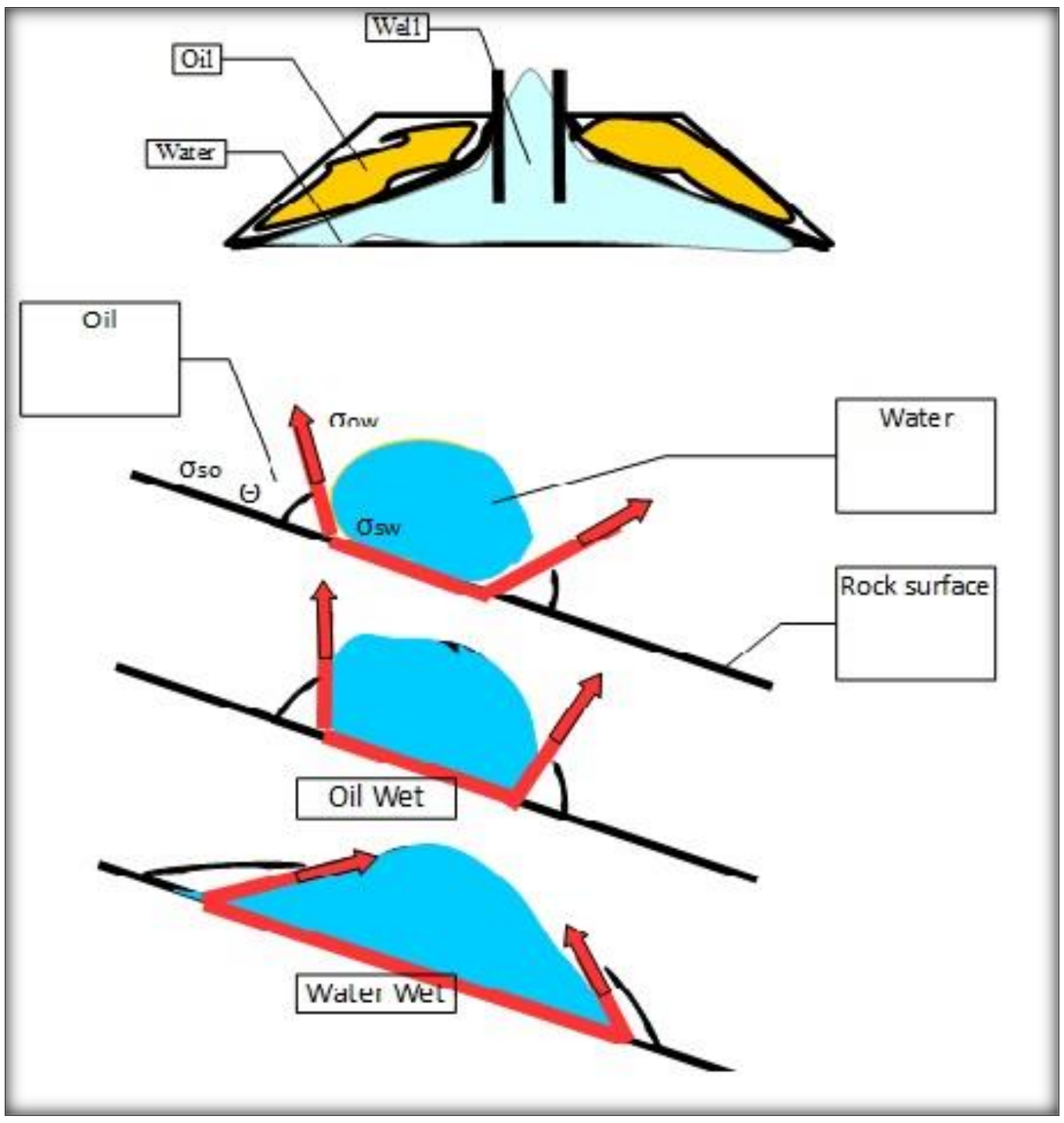

**Figure 2:** Schematic of the contact angle **Θ**, via water phase